\documentclass[12pt]{iopart}

\usepackage[left,modulo]{lineno}
\usepackage{iopams}
\usepackage{graphicx}
\usepackage{color}
\usepackage{ulem}

\begin{document}

\title{Stability comparison of two absolute gravimeters: optical versus atomic interferometers }

\author{P. Gillot$^1$,  O. Francis$^2$, A. Landragin$^1$,  F. Pereira Dos Santos$^1$ and S. Merlet$^1$}

\address{$^1$LNE-SYRTE, Observatoire de Paris, LNE, CNRS, UPMC,
61 avenue de l'Observatoire, 75014 Paris, France.}

\address{$^2$Faculty of Science, Technology and Communication, University of Luxembourg (UL), L-1359, Luxembourg}

\ead{sebastien.merlet@obspm.fr}

\begin{abstract}

We report the direct comparison between the stabilities of two mobile absolute gravimeters of different technology: the LNE-SYRTE Cold Atom Gravimeter and FG5X\#216 of the Universit\'e du Luxembourg. These instruments rely on two different principles of operation: atomic and optical interferometry. The comparison took place in the Walferdange Underground Laboratory for Geodynamics in Luxembourg, at the beginning of the last International Comparison of Absolute Gravimeters, ICAG-2013. We analyse a 2h10 duration common measurement, and find that the CAG shows better immunity with respect to changes in the level of vibration noise, as well as a slightly better short term stability.

\end{abstract}

\maketitle

\section{Introduction}\label{intro}

Absolute gravimeters measure test body free fall acceleration. The most used is the state-of-the-art commercial gravimeter FG5~\cite{Niebauer1995}. It measures the free fall of a corner cube with a Mach-Zehnder interferometer. Since the beginning of the 90's~\cite{Kasevich1992}, laboratories started to elaborate gravimeters using cold atoms as a test mass~\cite{Peters2001, LeGouet2008, Zhou2012}. This led to develop transportable instruments~\cite{Bodart2010, LouchetChauvet2011, Bidel2013, Hauth2013} coming out on participation to International Comparisons of Absolute Gravimeters (ICAG) as the LNE-SYRTE Cold Atom Gravimeter (CAG) does since 2009~\cite{Jiang2012, Francis2013}.

Usually, free fall corner cube users and in particular FG5 operators, record gravity by sets consisting of a number of drops (of the order of 100), that get repeated every hour. The repetition rate is usually of order of one drop every 10~s, in order to wait for the damping of the vibrations due to the carriage free fall and to preserve the device from mechanical wear. In \cite{Merlet2010}, one free fall per 30~s was chosen, leading to an Allan standard deviation about twice worse than if one drop per 10~s would have been chosen. On one hand, the FG5 dropping chamber~\cite{Niebauer2011} allows drops of 2~s which can improve notably the stability of FG5. On the other hand, alike the FG5, that uses a sophisticated super-spring system~\cite{Nelson1991}, various vibration rejection systems have been demonstrated and gradually improved over the last years to reject ground noise for atom sensors. They are based on the combination of a passive isolation and of a low noise seismometer \cite{LeGouet2008, Merlet2009}. Eventually, it turned into an active system~\cite{Hensley1998, Hauth2013, Zhou2012}, by using the signal of the seismometer to even better stabilize the position of the reference mirror. Using such an optimized active system, a stability of 4.2~$\mu$Gal\footnote{1~Gal~=~1~cm.s$^{-2}$, 1~$\mu$Gal~=~10$^{-8}$~m.s$^{-2}$} in 1~s measurement time was demonstrated in~\cite{Hu2013}. Using a passive system, or set directly on the ground, the CAG demonstrated a stability of 1~$\mu$Gal in 100~s measurement time interval~\cite{Farah2014}. To compare the stability performances of both technologies it is desirable to perform measurements at the same place and at the same time, under the influence of the same vibration noise. We took advantage of the last ICAG which took place in the Walferdange Underground Laboratory for Geodynamics (WULG) in Luxembourg at the end of 2013 to test the capabilities of the FG5X\#216 and CAG on a common view measurement.

\section{Measurements}\label{section1}

\begin{figure}[h!]
 \centering
 \noindent\includegraphics[width=250pc]{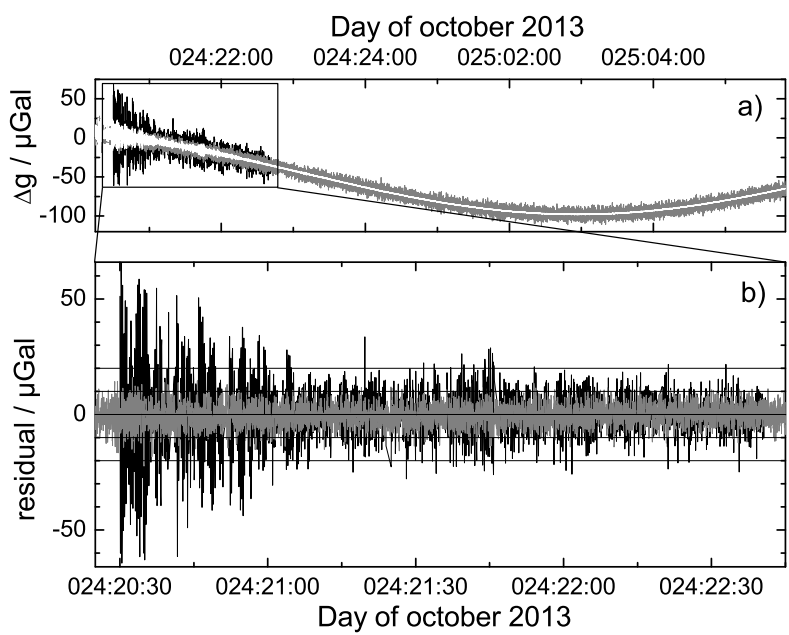}
 \caption{a) Earth's gravity variation during the night from the 24$^{th}$ to the 25$^{th}$ of october 2013 measured at Walferdange with FG5X$\#$216 in black and CAG in grey. The gravity variations observed with the superconducting gravimeter OSG-CT040 is also plotted in white. b) FG5X\#216 and CAG corresponding signals corrected for tides and atmospheric pressure effects, during the common view measurement.}
 \label{figg1}
 \end{figure}

\begin{figure}[h!]
 \centering
 \noindent\includegraphics[width=30pc]{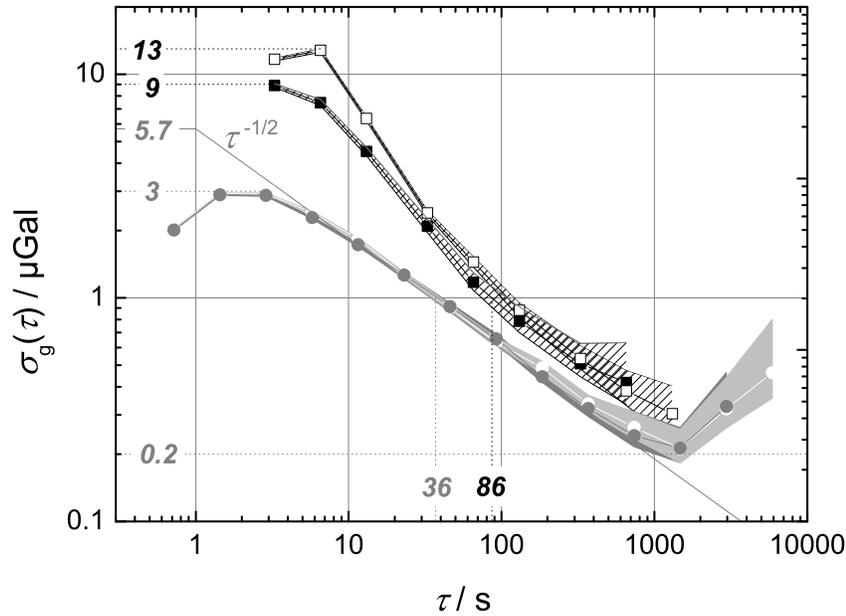}
 \caption{Allan standard deviation of the corrected gravity signals. FG5X\#216: open squares take into account all the drops and black squares exclude the earthquake. CAG: open circles take into account all the drops and grey circles exclude the earthquake. The $\tau^{-1/2}$ slope represents the averaging expected for white noise, the filed areas surrounding the Allan standard deviations dots represent the confidence intervals of the analyzes.} 
 \label{figallan}
 \end{figure}

Both gravimeters were installed on the platform B of the WULG~\cite{Francis2013}. The common view measurements were performed during the night between the 24$^{th}$ and the 25$^{th}$ of october 2013. The drop interval of the FG5X was chosen at 3~s close to his best capability of 2~s. We decided to use 3~s over 2 hours to spare the moving mechanical part of the instrument. The CAG measured continuously, all night long, using the protocol already used in~\cite{Farah2014} which is based on two interleaved integrations leading to a measurement time of 720~ms. Measurements are represented on figure~\ref{figg1}. It started at 20h15 for CAG and 15 minutes later for the FG5X. We realized only after the measurement was performed that the seismic noise was relatively high initially, due to an earthquake of magnitude 6.7 that occured in the East of South Sandwich Islands. This excess noise can be seen on the FG5X first half hour measurement as well as on the superconducting gravimeter OSG-CT040 that records gravity variation continuously a few meters only from platform B.  Usually, FG5 users compute the "drop scatter" (the standard deviation of a set) to characterize the dispersion of the measurements. Here the drop scatter of the FG5X first half hour measurement is 21.7~$\mu$Gal and 9.1~$\mu$Gal after. A zoom on the first hours of the gravity signals corrected from tides and atmospheric pressure effects (Figure~\ref{figg1} b) shows that the CAG is almost not affected by the seismic wave. The vibration rejection system~\cite{LeGouet2008, Merlet2009} is good enough to suppress the effect of the earthquake. This can also be seen on figure~\ref{figallan}.

In this paper, we choose to analyse the stabilities of the measurements using the Allan standard deviations~\cite{Allan1987} of the corrected gravity data (figure~\ref{figallan}). Two analyzes were performed for each gravimeter with and without the period during which the influence of the earthquake is significant. As we can guess from figure~\ref{figg1}, the short term stability of the FG5X is about 30~\% better when excluding the first half hour. After 200~s of measurement time the Allan standard deviation calculated with and without the earthquake noise are similar. The 1~$\mu$Gal level is obtained after 86~s of measurement and the Allan standard deviation continues to decrease down to 0.3~$\mu$Gal and maybe even better. However the FG5X measurements would have to be longer to perform this analyzis. For shorter averaging times, the Allan standard deviation decreases faster than a $\tau^{-1/2}$ slope. This behaviour is due to the averaging of the low frequency noise which is resonably well sampled by the FG5X. As a consequence the statistical error $se$ should not be estimated here using the standard formula $se=sd/\sqrt{N}$ where $N$ is the number of free falls. In contrast, the CAG stability is not affected by the seismic wave. We find that the Allan standard deviation for the whole CAG measurements, displayed with open circles on figure~\ref{figallan} is superimposed with the grey circles representing the Allan standard deviation calculated when excluding the earthquake. The initial bump on the Allan standard deviation is due to our measurement technique: the CAG signal is locked onto the gravity acceleration thanks to an integrator with a time constant of a few cycles~\cite{Merlet2009}. Then the Allan deviation decreases with a $\tau^{-1/2}$ slope up to 370~s (the 1~$\mu$Gal level is obtained after less than 36~s of measurement) and continues to decrease down to 0.2~$\mu$Gal. Such a long term stability had already been obtained in the LNE laboratory~\cite{Merlet2008}, by comparing the CAG with a superconducting gravimeter iGrav~\cite{iGravGWR}. Performing the FG5X measurement every 2~s instead of 3~s would only slightly affect the Allan standard deviation by shifting the curve to the left and would be still above the CAG results. This can be inferred from a previous study on the uncertainty of the FG5~\cite{VanCamp2005} showing that at high frequency ($10^{-5}$~Hz $\leqslant \nu \leqslant 10^{-1}$~Hz) the noise of a FG5 is white. The same study reveals that the noise increases at lower frequencies due to gravity changes linked to environmental fluctuations, that are not modeled. Both Allan variance curves of the CAG and FG5X will thus overlap for an integration time greater than one fourth day.
 
\section{Conclusion}

We compared the stabilites of two absolute gravimeters of different technology. Atom interferometry was already known for high cycling rate operation and the new FG5X for performing a free fall measurement every 3~s. During a quiet period the FG5X reaches a stability of 1~$\mu$Gal after 86~s averaging time while the CAG needs only 36~s, even for higher level of vibration noise. Considering the current level of accuracy of such gravimeters, of order of a few $\mu$Gal at best, a measurement time of only a few minutes is enough for the statistical uncertainty to be a negligible contribution to the combined uncertainty in the measurement. 

The possibility to perform continuous measurements with atom gravimeters at high cycling rates, and to reach stabilities of 0.2~$\mu$Gal in less than 2~000~s, now offer the opportunity to develop such instruments for permanent installation in geophysical observatories. Moreover, the sensitivity of atom gravimeters which scales as $T^2$ can be increased using taller vacuum chambers and larger time $T$ between the three interogating pulses~\cite{Borde2001}. As an example, the fountain configuration used in~\cite{Hu2013} allows to increase $T$ up to 300~ms, to be compared with the 80~ms we use in the CAG.

\ack
This research is carried on within the kNOW project, which acknowledges the financial support of the EMRP. The EMRP was jointly funded by the European Metrology Research Programme (EMRP) participating countries within the European Association of National Metrology Institutes (EURAMET) and the European Union. The CAG participation to ICAG-2013 was also supported by GPhys of Observatoire de Paris.

\end{document}